\documentclass[twocolumn,prl,aps,epsfig]{revtex4}
\usepackage[latin9]{inputenc}
\usepackage{amsmath}
\usepackage{amsthm}
\usepackage{amssymb}
\usepackage{esint}
\usepackage[unicode=true,
 bookmarks=false,
 breaklinks=false,pdfborder={0 0 1},backref=section,colorlinks=false]
 {hyperref}
\hypersetup{
 colorlinks,citecolor=red}

\makeatletter
\@ifundefined{textcolor}{}
{%
 \definecolor{BLACK}{gray}{0}
 \definecolor{WHITE}{gray}{1}
 \definecolor{RED}{rgb}{1,0,0}
 \definecolor{GREEN}{rgb}{0,1,0}
 \definecolor{BLUE}{rgb}{0,0,1}
 \definecolor{CYAN}{cmyk}{1,0,0,0}
 \definecolor{MAGENTA}{cmyk}{0,1,0,0}
 \definecolor{YELLOW}{cmyk}{0,0,1,0}
}

\usepackage{amsfonts}
\usepackage{amscd}\usepackage{amsthm}\usepackage{mathrsfs}
\usepackage{physics}
\usepackage{units}

\makeatother

\begin{document}
\title{Robust gate design for large ion crystals through excitation of local
phonon modes}
\author{L.-M. Duan\thanks{lmduan@tsinghua.edu.cn}}
\affiliation{Center for Quantum Information, Institute for Interdisciplinary Information
Sciences, Tsinghua University, Beijing 100084, PR China}
\begin{abstract}
We propose a scalable design of entangling quantum gates for large
ion crystals with the following desirable features: 1) The gate design
is universal and applicable for large ion crystals of arbitrary sizes;
2) The gate has no speed limitation
and can work outside of the Lamb-Dicke region; 3) The gate operates
by driving from either continuous-wave or pulsed laser beams; 4) The
gate is insensitive to slow variation of the laser optical phase and
works under a thermal state for the ions' motion; 5) The intrinsic
gate infidelity can be reduced to a level well below the threshold
for fault-tolerant quantum computation under realistic experimental
parameters. Different from the previous gate schemes, here we propose
a gate design based on driving of the local oscillation mode of the
ions instead of the collective normal modes and develop a formalism
based on the Heisenberg equations to deal with the many-body quantum
dynamics outside of the Lamb-Dicke region.
\end{abstract}
\maketitle
Trapped ions constitute one of the most promising systems for realization
of large-scale quantum computing \cite{1,2,3,4}. To scale up the trapped ion
quantum computer, several approaches have been considered, including
the ion shuttling approach based on the QCCD (quantum charge-coupled
device) architecture \cite{5,6}, the quantum network approach based on
the photon entanglement links of ions in different traps \cite{7,8,9}, and
the direct approach based on entangling gates in large ion crystals
of 1D (one-dimensional) \cite{10}, 2D \cite{11}, or 3D \cite{12} geometry.
The latter approach, when it works, would be most convenient and cost-saving
for experiments. Even with the first two approaches in mind for the
ultimate scaling, it is always cost-effective for large-scale computing
to have as many ions as possible in local ion crystals. With use of
the cryogenic traps, one can stably control any large ion crystals
with negligible influence from the background gas collisions \cite{12a}. The
major challenge is then the design of robust entangling
gates in large ion crystals of arbitrary sizes.

The entangling gates play a central role for implementation of quantum computing. For
trapped ion systems, the original Cirac-Zoller gate assumes the ground
state cooling of the ions \cite{1}. The Molmer-Sorensen gate \cite{13,14} and the
phase gate \cite{15,16}, which are widely used in experiments, alleviate
this requirement and replace it with the Lamb-Dicke condition. These
gates typically still assume sideband addressing of individual canonical
normal modes during the gates. For large ion crystals, in particular
for more scalable gates based on excitation of the transverse phonon
modes \cite{17}, it is challenging to satisfy the sideband addressing
condition, and the gate scheme based on segmentally modulated laser
pulses was introduced in Ref. \cite{17,18}, which removes the sideband addressing
condition and finds wide applications in recent experiments for gates
based on excitation of multiple phonon modes \cite{19,20,21,22,23,24,25,26}. These gates can
be extended to 2D and 3D ion crystal architectures by incorporating
the effects of micromotion in the gate design \cite{11,12}, and the segmental
modulation can be applied on either laser amplitude \cite{17,18,19,20,21,22}, frequency
detuning \cite{23,24}, or phase \cite{25,26}. One complexity for these segmentally
modulated gates is that the optimal gate parameters have to be calculated
based on the detailed experimental configuration, including the number
of ions in the crystal, the equilibrium positions of every ions, and
the normal mode spectrum. An important constraint for all these gates
is the requirement of the Lamb-Dicke condition, assuming the conditional
position shift of each ion much less than the laser wavelength. Breach
of this condition is a major cause of the gate infidelity when
one increases the gate speed \cite{22}. Another paradigm for the gate design is
based on application of a number of discrete momentum kicks from a train of
ultra-short pi-pulses \cite{27,28}.
This approach does not require the Lamb-Dicke condition, however, the complexity and the
error accumulation in applying many pi-pulses from an ultrafast laser make
the experiment for this approach quite challenging \cite{29}, and the achieved
fidelity so far is still significantly lower compared with those of
the other approaches and the threshold for fault-tolerant quantum computing.

In this paper, we propose a robust and scalable gate design with the
following features: 1) the gate applies to any large ion crystals
without the requirement of sideband addressing, and at the same time
the gate design is universal and does not require detailed knowledge
of the ion number and configuration in the crystal and the spectrum of
the normal modes; 2) the gate has no speed limitation and can work outside
of the Lamb-Dicke region; 3) the gate can operate by driving
from conventional continuous-wave laser beams and is insensitive to
the optical phase fluctuation from driving laser beams coming in different
directions. Different from the previous gate designs, here we achieve
the conditional phase in entangling quantum gates based on driving
of the local phonon modes instead of the collective normal mode of
the ion crystal. We develop a formalism for the gate design based
on the Heisenberg equations to deal with nonlinear dynamics outside
of the Lamb-Dicke region and use the interaction picture to calculate
the gate infidelity from the quantum many-body dynamics. We find
that the intrinsic gate infidelity can be reduced to a level well
below the error threshold for fault-tolerant quantum computing under
reasonable experimental parameters for any large ion crystals in both 1D
and 2D configurations. The scheme also directly applies to a scalable 2D array
architecture of microtraps \cite{32,33,34,31}
to achieve entangling gates with fast enough gate speed
under large ion spacing and moderate laser power.

Now let us consider the gate design for a large ion crystal of
arbitrary size. A key concept in this design is the local phonon mode
for the ion oscillation, with its frequency defined as the oscillation
frequency $\omega_{l}$ of the target ion $l$ when all the other ions in
the crystal are fixed at their equilibrium positions. The value of
$\omega_{l}$ includes contribution from the trapping potential and
the Coulomb interaction from all the ions. We achieve the entangling
gate by driving two target ions in the many-ion crystal with the laser
frequency resonant (or near-resonant) with the local oscillation frequency
$\omega_{l}$.

The total Hamiltonian of the system can be written as $H=H_{0}+H_{1}$
with
\begin{equation}
H_{0}=\sum_{\mu}\left(\frac{p_{\mu}^{2}}{2m}+\frac{1}{2}m\omega_{\mu}^{2}x_{\mu}^{2}\right)+\sum_{i=1,2}\sigma_{i}V(x_{i})
\end{equation}
where $x_{\mu}$ denotes the coordinate operator (displacement from
the equilibrium position) and $p_{\mu}$ is the corresponding momentum
operator. The summation $\mu$ is over all the ions in the crystal,
which should be understood as $\mu=(\mu_{1,}\mu_{2}$) for a 2D crystal.
The summation $i$ is only over the two target ions $1,2$ on which
we want to apply the entangling gate through application of a spin-dependent
potential. The local oscillation frequency $\omega_{\mu}$ is defined
by $m\omega_{\mu}^{2}\equiv\frac{\partial^{2}}{\partial x_{\mu}^{2}}\left(V_{T}+V_{C}\right)$
for the $\mu$th ion in the crystal, where $V_{T}$ denotes the trapping
potential, and $V_{C}$ denotes the Coulomb energy with $V_{C}=\frac{1}{2}\sum_{\mu\neq\mu'}\frac{k_{c}e^{2}}{\left|x_{\mu}-x_{\mu'}\right|}$
($\left|x_{\mu}-x_{\mu'}\right|$ should be understood as distance
between the two vectors $x_{\mu}$ and $x_{\mu'}$ for the 2D case).
Note that if $V(x_{i})$ is a liner function of $x_{i}$, it reduces
to a spin-dependent force, which is the case when we apply the Lamb-Dicke
condition (expansion of $V(x_{i})$ to the first order of $x_{i}$) \cite{13,14,15}.
Here we consider a general spin-dependent potential which could be
outside of the Lamb-Dicke region (fast gates require large spin-dependent
position shifts of the target ions which drive them outside of the
Lamb-Dicke region). If the gate is driven by a pair of traveling-wave
laser beams, the potential $V(x_{i})$ has the form
\begin{equation}
V(x_{i})=2\hbar\left|\Omega\right|\cos\left[kx_{i}+\phi_{t}+\phi_{0}\right]
\end{equation}
where $k$ denotes the wavevector difference of the two beams along
the $x_{i}$ direction, $\phi_{t}$ is a time-dependent phase of $\Omega$
that will be controlled during the gate, and $\phi_{0}$ is the slowly
varying optical phase difference from laser beams coming in different
directions which is assumed to be fixed and unknown during each gate
but fluctuates from gate to gate. The spin operator $\sigma_{i}$
reduces to $\sigma_{iz}$ if we apply a spin-dependent ac-Stark shift
\cite{16} and to $\sigma_{ix}$ if we apply the phase-insensitive Raman
driving \cite{30}. The Hamiltonian $H_{1}$ has the form
\begin{equation}
H_{1}=-\frac{1}{2}\sum_{\mu\neq\mu'}m\omega_{\mu,\mu'}^{2}x_{\mu}x_{\mu'}
\end{equation}
where $m\omega_{\mu,\mu'}^{2}\equiv\frac{-\partial^{2}V_{C}}{\partial x_{\mu}\partial x_{\mu'}}$.

We solve the dynamics in the interaction picture. With respect to
$H_{0}$, the Heisenberg equations for $x_{i}$ and $p_{i}$are given
by$ $
\begin{eqnarray}
\dot{x}_{i} & = & \frac{p_{i}}{m},\;\dot{p}_{i}=-m\omega_{i}^{2}x_{i}-\sigma_{i}\frac{\partial V(x_{i})}{\partial x_{i}}
\end{eqnarray}
From the driving of the spin-dependent potential $\sigma_{i}V(x_{i})$,
the $i$th ion will follow a spin-dependent trajectory in the phase
space. We decompose the operators $x_{i},p_{i}$ as
\begin{equation}
x_{i}=x_{ic}+x_{iq},\;p_{i}=p_{ic}+p_{iq},
\end{equation}
where $x_{ic},p_{ic}$ denote the classical spin-dependent trajectory
(proportional to $\sigma_{i}$, but otherwise a classical function
with $x_{ic}(0)=p_{ic}=0$ at the initial time), and $x_{iq},p_{iq}$
with $\left[x_{iq},p_{iq}\right]=i\hbar$ denote the small quantum
fluctuation around the spin-dependent trajectory. With $V(x_{i})$
given by Eq. (3), we can expand
\begin{align}
-\frac{\partial V(x_{i})}{\partial x_{i}} & =2\hbar\left|\Omega\right|k\sin\left(kx_{i}+\phi_{t}+\phi_{0}\right)\nonumber \\
 & \simeq2\hbar\left|\Omega\right|k\left[\sin\left(kx_{ic}+\phi_{t}+\phi_{0}\right)\right.\label{6}\\
 & \left.+kx_{iq}\cos\left(kx_{ic}+\phi_{t}+\phi_{0}\right)\right]\nonumber
\end{align}
where $kx_{ic}$ may not be small under large driving and thus we
keep the exact function to all the orders of $kx_{ic}$. However,
$kx_{iq}=\eta_{i}(a_{i}+a_{i}^{\dagger})$ with the Lamb-Dicke parameter
$\eta_{i}\equiv k\sqrt{\hbar/\left(2m\omega_{i}\right)}$ representing
the small quantum fluctuation around the spin-dependent trajectory,
which is determined by the initial thermal fluctuation and is small
under a small Lamb-Dicke parameter $\eta_{i}$. In Eq. (7), we expand
$\frac{\partial V(x_{i})}{\partial x_{i}}$ to the linear order of
$kx_{iq}$ (this is equivalent to expansion of $V(x_{i})$ to the
quadratic order of $kx_{iq}$, one order higher than the conventional
spin-dependent force approximation already). Later, we will consider
expansion of $V(x_{i})$ to even higher orders of $kx_{iq}$ and show
that they give only small contribution to the gate infidelity under
typical experimental parameters. With the above expansion, the Heisenberg
equation (4) becomes
\begin{alignat}{1}
\dot{x}_{ic} & =\frac{p_{ic}}{m},\nonumber \\
\dot{p}_{ic} & =-m\omega_{i}^{2}x_{ic}+2\hbar\left|\Omega\right|k\sigma_{i}\sin\left(kx_{ic}+\phi_{t}+\phi_{0}\right),\label{eq:8}
\end{alignat}
and
\begin{alignat}{1}
\dot{x}_{iq} & =\frac{p_{iq}}{m},\\
\dot{p}_{iq} & =-m\omega_{i}^{2}x_{iq}+4m\omega_{i}\eta_{i}^{2}\left|\Omega\right|\sigma_{i}x_{iq}\cos\left(kx_{ic}+\phi_{t}+\phi_{0}\right).\nonumber
\end{alignat}

Equation (7) is a set of nonlinear differential equations for the
classical variables $x_{ic},p_{ic}$ which can be solved exactly (numerically)
to give the spin-dependent trajectory. The spin operator $\sigma_{i}$
takes the eigenvalue $\pm1$ under the component eigenstates $\left|0\right\rangle ,\left|1\right\rangle $.
Equation (8) is a set of linear differential equations for the operators
$x_{iq},p_{iq}$, which can also be solved exactly with knowledge
of $x_{ic}$. Typically, the amplitude $4\eta_{i}^{2}\left|\Omega\right|\ll\omega_{i}$
, so the second term in Eq. (8) representing the spin-dependent modulation
of the oscillation frequency is only a small perturbation, and its
effect can be included into the Hamiltonian $H_{1}$ instead of $H_{0}$
by the following transformation
\begin{flalign}
H_{0} & \rightarrow H_{0}^{t}=H_{0}+\sum_{i=1,2}H_{i}^{p},\nonumber \\
H_{1} & \rightarrow H_{1}^{t}=H_{1}-\sum_{i=1,2}H_{i}^{p},
\end{flalign}
where $H_{i}^{p}\equiv2m\omega_{i}\eta_{i}^{2}\left|\Omega\right|\sigma_{i}x_{iq}^{2}\cos\phi_{xt}$
and $\phi_{xt}\equiv kx_{ic}+\phi_{t}+\phi_{0}$. Under this transformation,
the second term in Eq. (8) is gone and therefore the motion of $x_{iq},p_{iq}$
is spin independent under the transformed Hamiltonian $H_{0}^{t}.$
We choose a basic time interval $\tau$ for the gate so that the solution
of $x_{ic}(\tau),p_{ic}(\tau)$ from Eq. (7) is independent of the
value of $\sigma_{i}$ (a convenient choice is the solution with $x_{ic}(\tau)=p_{ic}(\tau)=0$).
In this case, the motional dynamics from $H_{0}$ gets decoupled from
the spin state at $\tau$. The unitary operator $U_{0}(\tau)\equiv T\exp\left(-i\intop_{0}^{\tau}H_{0}^{t}(t)dt/\hbar\right)$,
where $T\cdots$ denotes time-ordered integration, therefore is independent
of the spin state of the ions. The interaction Hamiltonian is then
given by
\begin{alignat}{1}
H_{I}(t) & =U_{0}^{\dagger}(t)H_{1}^{t}(x_{\mu})U_{0}(t)\nonumber \\
 & =H_{1}^{t}(U_{0}^{\dagger}(t)x_{\mu}U_{0}(t))\nonumber \\
 & =-\sum_{\mu\neq\mu'}(m/2)\omega_{\mu,\mu'}^{2}x_{\mu}\left(t\right)x_{\mu'}\left(t\right)\\
 & -\sum_{i=1,2}2\eta_{i}^{2}\sigma_{i}m\omega_{i}\left|\Omega\right|x_{iq}^{2}\left(t\right)\cos\phi_{xt},\nonumber
\end{alignat}
where $x_{\mu}\left(t\right),x_{iq}\left(t\right)$ represent the
solution from the corresponding Heisenberg equations under the Hamiltonian
$H_{0}$. With $x_{\mu}\left(t\right)=\sqrt{\hbar/(2m\omega_{\mu})}(a_{\mu}+a_{\mu}^{\dagger})$
for $\text{\ensuremath{\mu\neq1,2}}$ and $x_{\mu}\left(t\right)=x_{\mu c}(t)+x_{\mu q}(t)=\sqrt{\hbar/(2m\omega_{\mu})}(\alpha_{\mu c}+a_{\mu}+\alpha_{\mu c}^{*}+a_{\mu}^{\dagger})$
for $\text{\ensuremath{\mu=1,2}}$, the solution from $H_{0}$ has
the form
\begin{align}
a_{\mu}(t) & =a_{\mu}(0)e^{-i\omega_{\mu}t}.
\end{align}
In term of $\alpha_{\mu c}=\sqrt{m\omega_{\mu}/(2\hbar)}x_{\mu c}+i\sqrt{1/(2m\hbar\omega_{\mu})}p_{\mu c}$,
$(\mu=1,2)$, the equation (7) takes the form
\begin{align}
\dot{\alpha}_{\mu c} & =-i\omega_{\mu}\alpha_{\mu c}\nonumber \\
 & +i(2\eta_{\mu}\left|\Omega\right|\sigma_{\mu})\sin\left[\eta_{\mu}(\alpha_{\mu c}+\alpha_{\mu c}^{*})+\phi_{t}+\phi_{0}\right]
\end{align}
with the initial condition $\alpha_{\mu c}(0)=0$. We take a constant
amplitude $\left|\Omega\right|$ for the Raman laser beams and only
tune the relative phase $\phi(t)$ between the beams to satisfy
the condition $\alpha_{\mu c}(\tau)=0$.

In the Hamiltonian $H_{I}(t)$, the term $m\omega_{1,2}^{2}x_{1c}\left(t\right)x_{2c}\left(t\right)/2$
only depends on the spin operators $\sigma_{1},\sigma_{2}$ and does
not couple to the motional modes. The integration of this term over
time gives the desired entangling gate on the ion spin qubits. The
other terms in $H_{I}(t)$ represent remaining spin-motion coupling,
which contribute to the gate infidelity. The rate for these spin-motion
coupling terms is of the order
\begin{equation}
m\omega_{\mu,\mu'}^{2}/(2m\sqrt{\omega_{\mu}\omega_{\mu'}})=k_{c}e^{2}/\left(md^{3}\omega_{\mu}\right)\equiv\omega_{I}.
\end{equation}
where we have taken $\mu,\mu'$ as the nearest neighbor (the one with
the highest interaction rate) in the lattice with lattice distance
$d$ and the local oscillation frequency $\omega_{\mu}=\omega_{\mu'}$.
The rate $\omega_{I}$ is a basic quantity that characterizes the
interaction rate in the ion lattice. We can also define a quantity
$t_{p}\equiv\sqrt{md^{3}/\left(k_{c}e^{2}\right)}=d/v_{p}$ to characterize
the phonon propagation time to the neighboring lattice site, where
$v_{p}=\sqrt{k_{c}e^{2}/\left(md\right)}$ characterizes the phonon
propagation speed in a large lattice \cite{31}. With this definition,
we have $\omega_{I}=1/(\omega_{\mu}t_{p}^{2})$.

We take the basic time interval $\tau$ to satisfy the condition $\omega_{I}\tau\ll1$
(the exact condition will be specified below when we derive the expression
for the gate infidelity). The evolution operator in the interaction
picture is expressed as
\begin{equation}
U_{I}(\tau)=Te^{-i\intop_{0}^{\tau}H_{I}(t)dt/\hbar}.
\end{equation}
Without loss of generality, we assume the target ions $1,2$ are at
the neighboring sites with $\eta_{1}=\eta_{2}=\eta$, $\omega_{1}=\omega_{2}=\omega$, and $\alpha_{1c}=\alpha_{2c}=\alpha_{\pm}$
, where $\alpha_{\pm}$ corresponds to
the solution of Eq. (12) with $\sigma_{\mu}=\pm1$. The conditional
phase term in $U_{I}(\tau)$ has the form $e^{i\Phi}$ with
\begin{flalign}
\Phi & =\varphi_{c}\sigma_{1}\sigma_{2}+\varphi_{s}\left(\sigma_{1}+\sigma_{2}\right)
\end{flalign}
where $\varphi_{c}=\omega_{I}\intop_{0}^{\tau}\left(\alpha_{R+}-\alpha_{R-}\right)^{2}dt$,
$\varphi_{s}=\omega_{I}\intop_{0}^{\tau}\left(\alpha_{R+}^{2}-\alpha_{R-}^{2}\right)dt$,
$\alpha_{R\pm}\equiv\text{Re}\alpha_{\pm}\left(t\right)$, and we
have dropped the spin-independent global phase $\frac{1}{2}\intop_{0}^{\tau}\omega_{I}\left(\alpha_{R+}+\alpha_{R-}\right)^{2}dt$
in $\Phi$.

For fast gates with $\eta\alpha\sim1$, the nonlinear equation
(12) can be solved numerically. There is a convenient choice
of $\phi_t$ to satisfy the condition $\alpha(\tau)=0$. We take
$\phi_t=\phi(t-\tau/2)$ to be an even function of $t-\tau/2$, and the equation
(7), which is equivalent to Eq. (12), has a solution with $x_{ic}(t-\tau/2)$=$x_{ic}(\tau/2-t)$
(even) and $p_{ic}(t-\tau/2)=-p_{ic}(\tau/2-t)$ (odd). We can
divide the duration $\tau/2$ into several time segments, and
for each time segment$j$, we take $\phi_{j}(t)=\pm\omega t+\phi_{j0}$
with appropriate $\phi_{j0}$ so that we have $\text{Im\ensuremath{\alpha(\tau/2)\propto p_{ic}(t=\tau/2)=0}}$.
Due to the symmetry, it is then obvious $\alpha(\tau)=\alpha(0)=0$.

When the system is in the Lamb-Dicke region with $\eta\alpha\ll1$, we
can derive an analytic expression for the solution $\alpha(t)$.
We take $\phi_t=\omega t$ when $0\leq t\leq\tau/2$ and $\phi_t=\omega t+\pi$
when $\tau/2 < t\leq\tau$, the solution $\alpha(t)$ is
given by $\alpha(t)=\eta\left|\Omega\right|\sigma e^{-i\omega t}\left[e^{i(\omega t+\phi_{0})}\sin(\omega t)-\omega t e^{-i\phi_{0}}\right]/\omega\:$
for $0\leq t\leq\tau/2$, and $\alpha(t)=\eta\left|\Omega\right|\sigma e^{-i\omega t}[-e^{i(\omega t+\omega\tau/2+\phi_{0})}\sin\omega(t-\tau/2)+e^{i(\omega\tau/2+\phi_{0})}\sin(\omega\tau/2)+\omega(t-\tau)e^{-i\phi_{0}}/\omega]\text{ for \ensuremath{\tau}/2\ensuremath{\leq}t\ensuremath{\leq\tau}}$.
We have $\alpha(\tau)=0$ when $\omega\tau=2K\pi$, where $K$ is an
integer. The conditional phase shift in this case is given by
\begin{equation}
\Phi=\eta^{2}\omega_{I}\tau\left|\Omega\right|^{2}/(6\omega^{2})\left[\omega^{2}\tau^{2}+36\cos^{2}\phi_{0}-6\right]\sigma_{1}\sigma_{2}.
\end{equation}

Apart from the above conditional phase shift term $\Phi$, the other
terms in the interaction Hamiltonian $H_{I}$, denoted as $H_{Ir}$,
generate residue spin-motion entanglement at time $\tau$ which
contributes to the gate infidelity. The evolution operator in the interaction
picture can be expressed as
\begin{equation}
U_{I}(\tau)=e^{i\Phi}Te^{-i\intop_{0}^{\tau}H_{Ir}dt/\hbar}=e^{i(\Phi+A_{+}+A_{-})},
\end{equation}
where $A_{\pm}$ denotes the part of the generator that doesn't (does)
flip a sign when we flip the sign of the spin operator $\sigma_{i}$
($i=1,2)$. To suppress the spin-motion entanglement after the gate,
similar to the idea of dynamical decoupling, we compose $2^{n}$ ($n=1,2,3,\cdots)$
segments of the basic time step $\tau$. For each segment, we control
$\phi(t)$ to be identical except for an additional phase $\phi_{aj}$
($j=1,2,\cdots,2^{n}$) of $0$ or $\pi$ (a phase $\pi$ corresponds
to a sign flip of $\sigma_{i}$). With $n=1,2,3$, we call the resultant
scheme the $\phi2$, $\phi4$, $\phi8$ protocol, with the explicit
sequence of phase $\phi_{aj}$ for each segment: $\phi2:\phi_{aj}=[0,\pi]$,
$\phi4:\phi_{aj}=[0,\pi,\pi,0]$, $\phi8:\phi_{aj}=[0,\pi,\pi,0,\pi,0,0,\pi]$.
Denote the corresponding evolution operator for the $\phi2$, $\phi4$,
$\phi8$ protocol by $U_{I}(2\tau)=e^{i(2\Phi+A_{2})}$, $U_{I}(4\tau)=e^{i(4\Phi+A_{4})}$, $U_{I}(8\tau)=e^{i(8\Phi+A_{8})}$,
respectively. The explicit expressions for $A_{2}$, $A_{4}$, $A_{8}$
can be derived using the Baker-Hausdorff formula and are given in
the supplementary materials. We calculate the gate infidelity with
the resultant $U_{I}$. Using the $\phi8$ protocol as an example,
as derived in the supplementary materials, the gate infidelity $\delta F$
is given by

\begin{equation}
\delta F\simeq\omega_{I}\tau\left(\eta\left|\Omega\right|\tau\right)^{2}(2n_{c}\omega_{I}\tau)^{7}\left(2\overline{n}+1\right),
\end{equation}
where $\overline{n}$ denotes the mean thermal phonon number of the local mode and $n_{c}$
is a dimensionless parameter roughly estimated by the lattice coordination
number with $n_{c}\simeq2.0$ ($5.6$) for 1D (2D) ion lattice.

The conditional phase in Eq. (16) acquired by this gate has a dependence
on the unknown optical phase $\phi_{0}$ through its second term (coming
from the oscillating terms usually neglected by the rotating-wave
approximation). Such a dependence becomes negligible for conventional
slow gates with $\omega\tau>>2\pi$, however, its magnitude is comparable
with the first term when $\omega\tau=2\pi$. To remove the dependence
on the phase $\phi_{0}$, we combine two $\phi8$ sequence in succession,
while adding a phase of $\pi$/2 to $\phi(t)$ for the second $\phi8$
sequence. The conditional phase acquired by the two combined $\phi8$
sequences is then
\begin{equation}
\Phi=(8/3)\omega_{I}\tau\left(\eta\left|\Omega\right|\tau\right)^{2}\left[1+12/(\omega\tau)^{2}\right]\sigma_{1}\sigma_{2}.
\end{equation}
To realize a controlled phase flip (CPF) gate, we need $\Phi=\pi\sigma_{1}\sigma_{2}/4$,
and this condition sets the time $\tau$ (or with a fixed $\tau$
sets the magnitude of the Raman Rabi frequency $\left|\Omega\right|$).
In this case, the intrinsic gate infidelity is twice the value given
in Eq. (18).

Let us estimate the gate performance under some experimental
parameters. For ions with the Lamb-Dicke parameter $\eta\approx0.05$
and the local oscillation frequency $\omega=2\pi\times3$
MHz (typical for driving transverse modes of 1D or 2D ion
crystals), we have $\omega_{I}\simeq2\pi\times10$ ($3.6$)
KHz under an ion spacing of $d=8.8$ ($12.4$) $\mu$m. Let us take
$2n_{c}\omega_{I}\tau\simeq1/4$ ($\omega\tau=6\pi$ with $n_{c}\simeq2.0$
($5.6$) for 1D (2D) ion crystals). To realize a CPF gate, the required
$\left|\Omega\right|\simeq2\pi\times6.8$ ($11.5$) MHz for 1D (2D) crystals. The intrinsic
gate infidelity is $\delta F=1.0\times10^{-4}$ with
the mean thermal phonon number $\overline{n}\sim 1$. The gate time
in this case is $T_g=16\tau=16$ $\mu$s. For 2D arrays of microtraps with large ion
spacing $d\sim50$ $\mu$m \cite{32,33,34,31}, we can get the gate time $T_g=192$ $\mu$s and the intrinsic gate infidelity $\delta F=0.92\times10^{-7}$ by choosing $\omega\tau=144\pi$ and
$\left|\Omega\right|\simeq2\pi\times 1.1$ MHz using a single $\phi8$ cycle.

Another contribution to the gate infidelity is from the higher-order
Lamb-Dicke expansion in $kx_{iq}=\eta(a_{i}e^{-i\omega t}+a_{i}^{\dagger}e^{i\omega t})$.
In Eq. (6), we have included the expansion to the first two orders
in $kx_{iq}$ and all the orders in $kx_{ic}$. In the supplementary
materials, we show that the higher-order expansion terms in $\sin\left(kx_{i}+\phi_{t}+\phi_{0}\right)$
of Eq. (6) contribute to the gate infidelity by the form $\delta F\simeq\frac{\pi^{2}}{2}\eta^{4}\left(\bar{n}+1/2\right)^{2}$$\simeq0.38\times10^{-4}$
under $\overline{n}\sim 1$. For 2D ion crystals, the ions
also have micromotion around their equilibrium positions. As shown
explicitly in Ref. \cite{11}, the micromotion has well defined dynamics
which does not give additional gate infidelity but leads to a renormalization
of the effective Raman Rabi frequency $\left|\Omega\right|$ by an
average over the Gaussian beam profile in the 2D plane (assuming the wavevector difference of the
Raman beams is perpendicular to the ion lattice to drive the transverse
modes). After this correction of the effective magnitude $\left|\Omega\right|$,
the above formalism remains unchanged.

In summary, we have developed an approach for designing robust and
scalable entangling gates for ions in arbitrarily large lattices based
on excitation of the local phonon modes. The scheme has a number of
desirable features, removes some key limitations in the current approach,
and may have wide applications in future experiments as one scales
up the ion trap quantum computers.

I thank Y.-K. Wu for discussions. This work was supported by the Tsinghua University Initiative Scientific Research Program and the Ministry of Education of China through its fund to the IIIS.

\end{document}